\documentclass[epj]{svjour}

\newcommand{\be}{\begin{equation}}
\newcommand{\ee}{\end{equation}}

\newcommand{\bea}{\begin{eqnarray}}
\newcommand{\eea}{\end{eqnarray}}
\newcommand{\beann}{\begin{eqnarray*}}
\newcommand{\eeann}{\end{eqnarray*}}
\newcommand{\nn}{\nonumber}
\newcommand{\ba}{\begin{array}}
\newcommand{\ea}{\end{array}}

\newcommand{\LR}{\Leftrightarrow}
\newcommand{\mybar}[1]%
        {\kern 0.6pt\overline{\kern -0.6pt#1\kern -0.6pt}\kern 0.6pt}

\newcommand{\Tr}{{\rm Tr}\,}

\newcommand{\R}{\mathbb{R}}
\newcommand{\C}{\mathbb{C}}

\newcommand{\del}{\partial}

\newcommand{\lambdat}{\tilde{\lambda}}
\newcommand{\Phib}{\bar{\Phi}}
\newcommand{\nv}{{\bf n}}

\newcommand{\iv}{{\bf i}}
\newcommand{\jv}{{\bf j}}
\newcommand{\N}{\mathbb{N}}

\usepackage{amssymb,fontenc,times,mathptmx,graphicx}
\usepackage{fancyhdr}

\def\mytitle{Lattice Formulations of
Two Dimensional Topological Field Theories} 
\def\myauthors{Tomohisa Takimi}  
\def\mytype{Parallel Talk}
\def\mysession{Theoretical Models}

\begin{document}
\title{A Lattice Formulation of
Two Dimensional Topological Field Theory}
\author{Kazutoshi Ohta \inst{1}
\thanks{\emph{kohta@phys.tohoku.ac.jp} }%
 \and
 Tomohisa Takimi \inst{2}
\thanks{\emph{ttakimi@riken.jp} }%
}                     
%
%
\institute{High Energy Theory Group
Department of Physics
Tohoku University
Sendai 980-8578, JAPAN
\and \textit{(presenter!)}
Theoretical Physics Laboratory, 
RIKEN, 2-1 Hirosawa, Wako, Saitama 351-0198, Japan}
%
\date{}
\abstract{We propose a non-perturbative criterion to investigate
whether supersymmetric lattice gauge theories
preserving partial SUSY can have the desired continuum limit or not.
Since the target continuum theories of the lattice models
are extended supersymmetric gauge theories including the
topological field theory (TFT) as a special subsector,
the continuum limits of them should reproduce the properties
of the TFT.
Therefore, whether the property of the TFT
can be recovered at the continuum limit becomes a non-perturbative criterion.
Then we accept it as a criterion.
In this paper, among the topological properties,
we investigate the BRST cohomology on the 
two dimensional ${\mathcal N}=(4,4)$ CKKU lattice
model without moduli fixing mass term. 
We show that 
the BRST cohomology in the target continuum theory
cannot be realized from the 
BRST cohomology on the lattice.
From this result,
we obtain the possible implication that
the ${\mathcal N} = (4,4)$ CKKU model cannot recover
the target continuum theory if the non-perturbative 
effects are taken into account.
} 
\maketitle
\section{Introduction}
\label{intro}
Supersymmetry is one of the main subjects in the particle physics.
The supersymmetric gauge 
theories exhibits a variety of complex non-perturbative phenomena
which have been vigorously investigated.
For example, there are many analytic studies about the 
Seiberg-Witten theory~\cite{Seiberg:1994rs} and AdS/CFT 
duality~\cite{Aharony:1999ti}.
Such approaches to non-perturbative 
physics are 
based on the property of
duality.
We can learn much more from 
the numerical study using the lattice formulation,
which is more universal method, 
since the method would enable us
to calculate any 
observables.

In spite of the need for supersymmetric lattice model,
the construction of the lattice formulation applicable to the numerical study
is difficult. 
Since the supersymmetry including the infinitesimal translation in its 
algebra is broken on the lattice
which breaks the translational invariance,
the ordinary lattice formulations suffer from the fine-tuning problem.
Fine-tuning problem is the difficulty to recover the 
target continuum theory when the quantum effects are taken into account,
and it makes the computation time too huge to perform the 
practical numerical calculation.

To solve the fine-tuning problem, 
several lattice gauge theories which preserve partial 
supersymmetry on the lattice are 
proposed~\cite{Cohen:2003qw,Cohen:2003xe,Sugino:2003yb,Catterall:2004np}
recently.
They utilize the topological twisting which is picking up a
set of supersymmetry generators
which does not include the 
infinitesimal translation in its algebra.
In this way, partial supersymmetry can be preserved on the lattice.

It is very important to investigate
whether the models really solve the fine-tuning problem or not.
To do it, we should investigate whether they recover the 
target continuum theories or not.
In the perturbative level, such investigations have done 
well. (For example \cite{Onogi:2005cz})
But, on the other hand, 
there is not a sufficient study which takes the non-perturbative effects
into consideration.
Then we will non-perturbatively examine whether the 
models really solve the fine-tuning problem or not.

\section{The proposed non-perturbative criterion}
\label{Sec:2}
Note that the 
models can be regarded as the lattice regularization
of the topological field theory (TFT).
This is because
preserved supercharges on the lattice
are equivalent to the BRST charge in the TFT
obtained by the topological twisting.
The target continuum theories of these lattice models are
extended supersymmetric gauge theories including the TFT
as a special subsector.
Therefore 
the topological field theory in the continuum theory 
must be recovered in the 
continuum limits 
if the lattice models really recover the target continuum 
theories.

In this work, among the several properties of the TFT,
we investigate the behavior of the BRST cohomology~\cite{Ohta:2006qz}.
The BRST cohomology is defined with the 
vacuum expectation value $\langle \mathcal{O} \rangle$ of an
operator $\mathcal{O}$
vanishing under the operation of the BRST charge $Q$ (BRST closed) but 
not BRST exact.
The BRST exact is a quantity 
written by the $Q$-operation of a gauge invariant quantity.
We can obtain the $\langle \mathcal{O} \rangle$ exactly
by the semi-classical approximation since the quantity 
$\langle \mathcal{O} \rangle$
is independent of the gauge coupling due to the property of the Hilbert space
of the TFT.
Namely, $\langle \mathcal{O} \rangle$ can be regarded as 
one of the non-perturbative
quantities.
Therefore, by examining whether the BRST cohomology 
in the continuum theory 
can be recovered at the continuum limit or not,
we can non-perturbatively investigate whether a
lattice model can recover the continuum theory or not.

In this paper, we consider whether
${\mathcal  N} = (4,4)$
two dimensional CKKU model~\cite{Cohen:2003qw} really have the 
desired continuum limit or not.
To do it, we study the BRST cohomology on the lattice.
Then we compare the BRST cohomology on the lattice with the 
BRST cohomology in the continuum theory,
and we consider whether the BRST cohomology in the target theory really 
recovered in the continuum limit.
From this study, we consider whether the target theory is recovered in the 
continuum limit or not.

\section{The BRST cohomology in the target theory.}
\label{Sec:3}
To make a comparison between the BRST cohomology 
in the target continuum theory and the ones on the lattice,
we should explain the BRST cohomology in the target continuum theory.
The action of the continuum theory is written by the BRST exact form
as described at the eq.~(5.1) in the paper~\cite{Takimi:2007nn}.
The BRST transformation law of the continuum theory is given at the
eq.~(5.2) in the paper~\cite{Takimi:2007nn}.
Among the transformation laws in the eq.~(5.2),
we describe the transformation laws
\begin{eqnarray}
Q\phi &=& 0, \nonumber\\
Q v_\mu &=& \psi_\mu, \nonumber\\
Q \psi_\mu &=& i D_\mu \phi, \label{cont-BRST}
\end{eqnarray}
here, since we use these transformation laws to create the BRST cohomologies 
in the continuum theory.
In eq.~(\ref{cont-BRST}),
$v_\mu$ denotes the gauge field and the $\psi_\mu$ denotes the BRST
partner of the gauge field.

In the continuum theory, the BRST cohomologies are composed by
$\phi$, $v_\mu$ and $\psi_\mu$ at least.
To compose the BRST cohomologies by these fields, we can utilize the 
`decent relation' proposed by Witten~\cite{Witten:1988ze}.
Let us prepare the differential 0-form, 1-form and 2-form operator set
\bea
\mathcal{W}_0 &=& Tr \phi^2, \nonumber \\
\mathcal{W}_1 &=& Tr \phi \psi, \nonumber \\
\mathcal{W}_2 &=& Tr \phi (dv + v\wedge v) + \psi \wedge \psi,
\eea
where $\psi$ and $v$ are differential 1-form denoted by
$\psi = \psi_\mu dx^\mu$ and 
$v = v_\mu dx^\mu$.
Here $d$ denotes the exterior derivative.
The set satisfies the following the `decent relation' 
\bea
Q{\cal W}_0 &=& 0,  \\
Q{\cal W}_k &=& d {\cal W}_{k-1} \qquad (k =1,2). 
\label{decent}
\eea
Utilizing this property, the BRST closed operators ${\cal O}_k$
can be constructed by the integral of ${\cal W}_k$ $(k=1,2)$ over the 
$k$ dimensional homology cycle $\gamma_k$,
\be
{\cal O}_k \equiv \int_{\gamma_k} {\cal W}_k.
\ee
We can confirm that these operators are BRST closed
by the explicit calculation,
\be
Q{\cal O}_k = Q\int_{\gamma_k} {\cal W}_k =
\int_{\gamma_k} d{\cal W}_{k-1}= \int_{\partial \gamma_k} {\cal W}_{k-1}
= 0,
\ee
since any homology cycle does not have boundaries.
Also the ${\cal W}_0$ can be regarded as the BRST closed operators
due to the transformation law $Q \phi = 0$.

These ${\cal O}_k$ are BRST cohomologies although they are 
\textit{formally}
written by the BRST exact form,
\be
{\cal O}_1 = \int Q \Tr \phi v, \qquad
{\cal O}_2 = \int Q \Tr \psi \wedge v.\label{BRST}
\ee
The operators $\Tr \phi v$ and $\Tr \psi \wedge v$ are
not gauge invariant.
The BRST exact quantities are defined by the $Q$-operation of gauge invariant 
quantities.
Therefore these ${\cal O}_1$ and ${\cal O}_2$ are not BRST exact but
BRST closed quantities, namely the BRST cohomologies.
Here, please note that
the $Q$-operation changes the gauge transformation laws as
\bea
v_\mu &\to& g^{-1} v_\mu g + g^{-1} \partial_\mu g, \label{gauge-trans}\\
Qv_\mu = \psi_\mu &\to& g^{-1} \psi_\mu g.
\eea
This property plays an important role to create the gauge invariant BRST
cohomology from the $Q$-operation of the gauge variant quantity.

\section{The BRST cohomology on the two dimensional ${\mathcal N} =(4,4)$
CKKU lattice model.}
\label{Sec:4}
Next, let us consider the BRST cohomology on the two dimensional 
${\mathcal N} =(4,4)$
CKKU lattice model without moduli fixing mass term.
The action of the lattice model is written at eq.~(3.14)
in \cite{Cohen:2003qw}, and the preserved supercharges and 
their transformation laws are
given by eqs.~(3.2),(3.3),(3.5) and (3.6) in \cite{Cohen:2003qw}.
The action can be written by the equivalent BRST exact form described in 
eq.~(2.14),(2.15) 
in \cite{Takimi:2007nn},
where the BRST charge is given by the 
the linear combination of the original superchrages as
eq.~(2.11) in \cite{Takimi:2007nn}.
In fact, also the BRST exact action 
eq.~(3.6) in \cite{Ohta:2006qz} is completely equivalent to 
eq.~(2.11) in \cite{Takimi:2007nn}. 
One can check the equivalence by identifying the fields
as follows
\be
\begin{array}{ll}
X_\nv \LR \sqrt{2} z_{1,\nv}, & \lambda_\nv \LR \sqrt{2} \psi_{1,\nv},
\\
X^\dag_\nv \LR \sqrt{2} \mybar z_{1,\nv}, 
& \lambda^\dag_\nv \LR - \sqrt{2} \xi_{2,\nv},
\\
Y_\nv \LR \sqrt{2} z_{2,\nv},  & \lambdat_\nv \LR \sqrt{2} \psi_{2,\nv},
\\
Y^\dag_\nv \LR \sqrt{2} \mybar z_{2,\nv},  
& \lambdat^\dag_\nv \LR \sqrt{2} \xi_{1,\nv},
\\
\bar{\Phi}_\nv 
\LR \sqrt{2} z_{3,\nv},  &  \eta_\nv \LR \sqrt{2} 
(\psi_{3,\nv} - \lambda_{\nv}), 
\\
\chi^\C_\nv \LR \sqrt{2} \chi_{\nv}, 
& \chi^{\C\dag}_\nv \LR \sqrt{2} \xi_{3,\nv}, 
\\
H^\C_\nv \LR \sqrt{2} \tilde{\mybar G}_{\nv},  
& H^{\C\dag}_\nv \LR \sqrt{2} \tilde{G}_{\nv}, 
\\
\chi^\R_\nv \LR -i \sqrt{2}(\psi_{3,\nv} + \lambda_{\nv}), 
& H^\R_\nv \LR -\tilde{d}_\nv,
\\
\Phi_\nv \LR \sqrt{2} \mybar z_{3,\nv}.&
\end{array}
\label{correspo-catte}
\ee
In this paper, we use the BRST exact form eq.~(3.7) in \cite{Ohta:2006qz}
of the CKKU lattice action, 
\bea
S &=& Q \Xi \nn \\
\Xi
&=& \Tr\Bigg[\frac{1}{4}\eta_{ {\bf n}}[\Phi_{ {\bf n}},\Phib_{ {\bf n}}]
+\vec{\chi}_\nv\cdot
(\vec{H}_\nv-i\vec{\cal E}_\nv) \nn
\\
&&+\frac{1}{2}
\biggl\{
\lambda_\nv(X_{\nv}^\dag\Phib_\nv-\Phib_{\nv+\iv}X_{\nv}^\dag)\nn\\
&&\qquad
+\lambda_{\nv-\iv}^\dag(X_{\nv-\iv}\Phib_\nv-\Phib_{\nv-\iv}X_{\nv-\iv})\nn\\
&&\qquad
+\lambdat_\nv(Y_{\nv}^\dag\Phib_\nv-\Phib_{\nv+\jv}Y_{\nv}^\dag)
\nn\\
&&\qquad
+\lambdat_{\nv-\jv}^\dag(Y_{\nv-\jv}\Phib_\nv-\Phib_{\nv-\jv}Y_{\nv-\jv})
\biggr\}\Bigg], \nn\\
\\
{\cal E}^\R_\nv &=& -(X_\nv X_\nv^\dag
-X_{\nv-\iv}^\dag X_{\nv-\iv}
+Y_\nv Y_{\nv}^\dag
-Y_{\nv-\jv}^\dag Y_{\nv-\jv}),\nn\\
{\cal E}^\C_\nv&=&2i(X_\nv Y_{\nv+\iv}-
Y_\nv X_{\nv+\jv}).\nn
\eea
In the tree level,
the continuum limit of the eq.~(3.7) in \cite{Ohta:2006qz}
becomes the topological field theory action eq.~(3.11)
(or eq.~(5.1) in \cite{Takimi:2007nn}), which is equivalent to the 
two dimensional ${\cal N} =(4,4)$ super Yang-Mills theory.
In the continuum limit, the lattice field variable $\Phi$ becomes
the field $\phi$ in the continuum theory, and the gauge fields $v_\mu$
come from the bosonic link fields $X,X^\dag,Y,Y^\dag$.
The BRST partner of the gauge fields $\psi_\mu$ come from the 
fermionic link field, $\lambda,\lambda^\dag,\lambdat,\lambdat^\dag$.
For later use, we distinguish the degree of freedom as the two part
$\{ \Phi_\nv \}$, which is the set composed only by the field $\Phi$,
and the set ${\cal \vec{A}}_\nv$ which is composed by the other fields.

The BRST transformation laws are given in eq.~(3.7) in \cite{Ohta:2006qz},
\be
\begin{array}{ll}
QX_\nv=\lambda_\nv, & Q\lambda_\nv =\Phi_\nv X_\nv - X_\nv\Phi_{\nv+\iv},\\
QY_\nv=\lambdat_\nv, & Q\lambdat_\nv =\Phi_\nv Y_\nv - Y_\nv\Phi_{\nv+\jv},\\
QH^\R_\nv =[\Phi_\nv,\chi^\R_\nv], & Q\chi^\R_\nv = H^\R_\nv,\\
QH^\C_\nv =\Phi_\nv\chi^\C_\nv-\chi^\C_\nv\Phi_{\nv+\iv+\jv}, & Q\chi^\C_\nv = H^\C_\nv,\\
Q\Phib_\nv=\eta_\nv, & Q\eta_\nv=[\Phi_\nv,\Phib_\nv],\\
Q\Phi_\nv=0. & 
\end{array}\label{orb-BRST}
\ee
Note that this is a homogeneous transformation of ${\cal \vec{A}}_\nv$.
Therefore, the transformation can be written as the tangent vector
\bea
Q &=&\sum_\nv\Bigg[
\lambda_\nv\frac{\del}{\del X_\nv} + \lambda_\nv^\dag\frac{\del}{\del X_\nv^\dag}
+\lambdat_\nv\frac{\del}{\del Y_\nv}+\lambdat_\nv^\dag\frac{\del}{\del Y_\nv^\dag} \nn \\
&&\quad
+[\Phi_\nv,\chi^\R_\nv]\frac{\del}{\del H^\R_\nv}
+(\Phi_\nv\chi^\C_\nv-\chi^\C_\nv\Phi_{\nv+\iv+\jv})\frac{\del}{\del H^\C_\nv}
\nn \\
&&\quad
+(\Phi_\nv\chi^{\C\dag}_\nv-\chi^{\C\dag}_\nv\Phi_{\nv-\iv-\jv})
\frac{\del}{\del H^{\C\dag}_\nv}
+\eta_\nv\frac{\del}{\del \Phib_\nv}\nn \\
&&\quad
+(\Phi_\nv X_\nv - X_\nv \Phi_{\nv+\iv})\frac{\del}{\del \lambda_\nv}
+(\Phi_\nv X_\nv^\dag - X_\nv^\dag \Phi_{\nv-\iv})\frac{\del}{\del \lambda_\nv^\dag}\nn \\
&&\quad
+(\Phi_\nv Y_\nv - Y_\nv \Phi_{\nv+\jv})\frac{\del}{\del \lambdat_\nv}
+(\Phi_\nv Y_\nv^\dag - Y_\nv^\dag \Phi_{\nv-\jv})\frac{\del}{\del \lambdat_\nv^\dag} \nn \\
&&\qquad\quad
+\vec{H}_\nv\cdot \frac{\del}{\del \vec{\chi}_\nv}
+[\Phi_\nv,\Phib_\nv]\frac{\del}{\del \eta_\nv}
\Bigg].
\label{tangent-lat}
\eea
From this property, 
if we introduce another fermionic operator written by the tangent vector
\bea
\tilde{Q} &=& \sum_{\nv}
X_\nv \frac{\partial}{\partial \lambda_\nv}
+X^\dag_\nv \frac{\partial}{\partial \lambda^\dag_\nv}
+Y_\nv \frac{\partial}{\partial \lambdat_\nv}
+Y^\dag_\nv \frac{\partial}{\partial \lambdat^\dag_\nv}
\nn \\
&& \qquad
+\bar{\Phi}_\nv \frac{\partial}{\partial \eta_\nv}
+\vec{\chi}_\nv \cdot \frac{\partial}{\partial \vec{H}_\nv}
,\label{another-Q}
\eea
we can construct the number operator $\hat{N}_{\cal A}$,
which count the number of fields in the set ${\cal \vec{A}}_\nv$,
by the anti-commutation relation, 
\bea 
\{Q, \tilde{Q} \} &=&  \sum_\nv X_\nv \frac{\partial}{\partial X_\nv}
+X^\dag_\nv \frac{\partial}{\partial X^\dag_\nv}
+Y_\nv \frac{\partial}{\partial Y_\nv}
+Y^\dag_\nv \frac{\partial}{\partial Y^\dag_\nv}
\nonumber\\
&&\quad +\lambda_\nv \frac{\partial}{\partial \lambda_\nv}
+\lambda^\dag_\nv \frac{\partial}{\partial \lambda^\dag_\nv}
+\lambdat_\nv \frac{\partial}{\partial \lambdat_\nv}
+\lambdat^\dag_\nv \frac{\partial}{\partial \lambdat^\dag_\nv}
\nonumber\\
&& \quad
+\bar{\Phi}_\nv \frac{\partial}{\partial \bar{\Phi}_\nv}
+\vec{H}_\nv\cdot \frac{\partial}{\partial \vec{H}_\nv}
+\eta_\nv \frac{\partial}{\partial \eta_\nv}
+\vec{\chi}_\nv\cdot \frac{\partial}{\partial \vec{\chi}_\nv}\nonumber\\
&=&\hat{N}_{\cal A}.
\label{number}
\eea
Please note that any function of the field variables can 
be written in terms of a sum of eigenfunction $h$ of $\hat{N}_{\cal A}$,
namely
\be
h = \sum_{n_{\cal A}= 0}^{\infty} h_{n_{\cal A}},
\qquad
\hat{N}_{\cal A} h_{n_{\cal A}} = n_{\cal A} h_{n_{\cal A}}, \quad 
n_{\cal A} \in \{0 \} \cup \N, \label{number-decomposition}
\ee
since any term in the function $h$ has definite number of fields
in the set $\vec{\cal A}_\nv$.
In addition to this homogeneous property of the BRST charge $Q$,
this $Q$ does not change the gauge transformation law opposite to the 
continuum theory case.
One can confirm it by checking that each field 
resides on the same link or site 
as its corresponding 
BRST partner
described in the 
right hand sides of the 
BRST transformation laws eq.~(3.7) in
\cite{Ohta:2006qz} respectively
(see also Fig.~1 in \cite{Ohta:2006qz}).

From these properties of BRST charges, 
we can see that BRST cohomology must be composed only 
by $\Phi$ on the lattice.
We will show it.
First, let us consider the BRST closed function $h_{c}$ satisfying 
$ Q h_{c} = 0$.
From the property eq.~(\ref{number-decomposition}),
also $h_{c}$ can be decomposed by the sum of eigenfunctions of the operator
$\hat{N}_{\cal A}$,
\be
h_{c} = \sum_{n_{\cal A}= 0}^{\infty} h_{c, n_{\cal A}}.
\ee
Since the BRST operator is homogeneous transformation which does not 
change the number of fields in $\vec{\cal A}_\nv$, 
the BRST operator $Q$ commutes with the number operator $\hat{N}_{\cal A}$,
namely
\be
[ Q, \hat{N}_{\cal A} ] = 0.
\ee
Then, if $Q h_{c} = 0$, each eigenfunction $h_{c,n_{\cal A}}$ 
composing the function $h_{c}$ must be BRST closed,
\be
Q h_{c} = 0 \LR Q h_{c,n_{\cal A}} = 0, \quad 
( n_{\cal A}^{\forall} \in \{0\} \cup \N).
\ee
The BRST closed eigenfunctions $h_{c,n_{\cal A}}$ with non-zero eigenvalue
$n_{\cal A} \ne 0$ can be formally written as the 
BRST exact form since 
\be
h_{c,n_{\cal A}} = n_{\cal A}^{-1} \hat{N}_{\cal A} h_{c,n_{\cal A}}
= n_{\cal A}^{-1} \{ Q, \tilde{Q} \} h_{c,n_{\cal A}}
= n_{\cal A}^{-1}  Q \tilde{Q}  h_{c,n_{\cal A}}.
\label{Q-exact-form}
\ee
Here the $Q$-operation does not change the gauge transformation law.
Then, in the 
eq.~(\ref{Q-exact-form}),
$\tilde{Q}  h_{c,n_{\cal A}}$ must be gauge invariant if the 
function $h_{c,n_{\cal A}}$ is a gauge invariant function.
Therefore, 
in the BRST closed function $h_{c}$, BRST closed non-zero eigenfunction
$h_{c, n_{\cal A}}$
must be BRST exact.
Finally, we can see that the 
only the zero eigenfunction $h_{c,0}$, which is the polynomial 
composed only by $\Phi$,
can be the BRST cohomology 
among the eigenfunctions.
This is the end of proof. 

The above situation stands for any lattice spacing.
This tells that the BRST cohomology must be composed only by $\Phi$
no matter how the lattice spacing is small, 
namely even in the continuum limit.
Therefore the BRST cohomology in the target continuum theory, 
which are composed not only by $\phi$ but also by gauge fields 
$v_\mu$ and their partners $\psi_\mu$, 
cannot be realized from the BRST cohomology on the lattice.
Finally, we obtain the possible implication 
that the ${\cal N} = (4,4)$ CKKU lattice model cannot realize the desired 
target continuum theory.

\subsection{A reason why the BRST cohomology cannot be realized 
on the lattice}
\label{subSec:4-1}
Among the BRST cohomologies in the target theory, 
the quantities composed by $v_\mu$ and $\psi_\mu$,
which are 1-form and 2-form operators 
${\cal O}_1$ and ${\cal O}_2$,
are defined by the inner product of the homology cycle and its dual cohomology
of the base manifold.
Such inner products are topological quantities
which are invariant under the infinitesimal transformation 
of the base manifold.
It is generally difficult to construct such a topological quantities on the 
lattice.
On the lattice, 
gauge symmetry are defined with the gauge parameter which are completely
independent of the parameter on the neighbor sites.
Such a property of the lattice gauge symmetry admits the singular gauge 
transformation which prevents from the realization of the topological 
quantities on the lattice.
Therefore we could not obtain the BRST cohomologies which are composed by 
$v_\mu$ and $\psi_\mu$.

Inhomogeneous term 
$g^{-1} \partial_\mu g$ in the eq.~(\ref{gauge-trans})
are removed from the gauge transformation law of the corresponding 
link gauge fields due to the property of lattice gauge symmetry.
By this property, $Q$ on the lattice
does not change the gauge transformation law.
Also it would be 
the reason why
it is impossibility to create the BRST cohomology on the lattice.

\section{Conclusion and discussion.}
\label{Sec:5}
In this paper,
we have investigated whether the supersymmetric lattice model, which is 
the two dimensional ${\cal N} =(4,4)$ CKKU supersymmetric lattice model,
really recovers the target theory or not through the examining whether
the property of the TFT are really recovered in the continuum limit or not.
As the first step,
we estimate the situation 
by the comparison 
between the BRST cohomology on the 
two dimensional ${\cal N} =(4,4)$ 
CKKU lattice and the BRST cohomology 
in its target continuum theory.
By this study, we have understood that the 
BRST cohomology in the target continuum theory cannot be realized
from the 
BRST cohomology on the lattice.
This implies that there is a possibility that the CKKU lattice model 
cannot realize the desired target continuum theory in the 
continuum limit.

Moreover, we consider the reason of the impossibility. 
The reason of the impossibility would be that 
the BRST cohomology is a topological quantity
defined by the inner product of the homology cycle and its dual cohomology.
Such a topological quantity is generally difficult to be realized on the
lattice since the gauge symmetry on the lattice admits the 
singular gauge transformation which prevents us from
defining the topological quantity on the lattice.
From this observation, we can guess
that also other models like~\cite{Sugino:2003yb,Catterall:2004np}
might be difficult to recover the desired target theories.
But, from this, we could obtain the valuable strategy
to develop the lattice formulation which can easily recover the 
desired target continuum theory, namely the formulation applicable to 
the numerical study.
We propose that we should apply the Admissibility 
condition~\cite{Luscher:1981zq} etc, which
enables to define the topological quantity like the chiral anomaly,
to define the BRST cohomology on the lattice and to recover the 
desired target theory.
\section*{Acknowledgments}
We would like to thank T.~Asakawa, S.~Catterall, 
P.~Damgaard,
H.~Fukaya,
A.~Hanany, I.~Kanamori, H.~Kawai, Y.~Kikukawa, T.~Kuroki,
S.~Matsuura, H.~Pfeiffer, Y.~Shibusa, F.~Sugino, H.~Suzuki, 
M.~Unsal
and N.~Yokoi
for useful discussions and comments.
K.O is supported in part by the 21st Century COE Program at Tohoku University.
Especially,
T.T is very grateful to
the Theoretical Physics Laboratory
The Institute of Physical and Chemical Research (RIKEN)
for supporting to take part in the SUSY conference.


\end{document}